\documentclass{elsart}
\usepackage{amssymb}
\usepackage{amsmath}\allowdisplaybreaks
\usepackage{graphicx}
\usepackage[normalem]{ulem}
\usepackage[dvips]{color}
\def\beq{\begin{eqnarray}}
\def\eeq{\end{eqnarray}}
\def\eps{\epsilon}

\begin{document}
\begin{frontmatter}

\title{Revisiting the boiling of primordial quark nuggets at nonzero chemical potential}
\author [XMU,KEY]{Ang Li\thanksref{info1}},
\author [XMU,KEY,KEY2]{Tong Liu\thanksref{info2}},
\author [RIK]{Philipp Gubler\thanksref{info3}},
\author [PKU]{and Ren-Xin Xu\thanksref{info4}}
\address[XMU]{Department of Astronomy and Institute of Theoretical Physics and Astrophysics, Xiamen University, Xiamen 361005, China} \address[KEY]{State Key Laboratory of Theoretical Physics, Institute of Theoretical Physics, Chinese Academy of Sciences, Beijing 100190, China}
\address[KEY2]{Key Laboratory for the Structure and Evolution of Celestial Objects, Chinese Academy of Sciences, Kunming, Yunnan 650011, China}
\address[RIK]{RIKEN Nishina Center, RIKEN, Wako 351-0198, Japan}
\address[PKU]{School of Physics and State Key Laboratory of Nuclear Physics and Technology, Peking University, Beijing 100871, China}
\thanks[info1]{liang@xmu.edu.cn}
\thanks[info2]{tongliu@xmu.edu.cn}
\thanks[info3]{pgubler@riken.jp}
\thanks[info4]{r.x.xu@pku.edu.cn}

\begin{abstract}

The boiling of possible quark nuggets during the quark-hadron phase transition of the Universe at nonzero chemical potential is revisited within the microscopic Brueckner-Hartree-Fock approach employed for the hadron phase, using two kinds of baryon interactions as fundamental inputs. To describe the deconfined phase of quark matter, we use a recently developed quark mass density-dependent model with a fully self-consistent thermodynamic treatment of confinement. We study the baryon number limit $A_{\rm boil}$ (above which boiling may be important) with three typical values for the confinement parameter $D$. It is firstly found that the baryon interaction with a softer equation of state for the hadron phase would only lead to a small increase of $A_{\rm boil}$. However, results depend sensitively on the confinement parameter in the quark model. Specifically, boiling might be important during the Universe cooling for a limited parameter range around $D^{1/2} = 170$ MeV, a value satisfying recent lattice QCD calculations of the vacuum chiral condensate, while for other choices of this parameter, boiling might not happen and cosmological quark nuggets of $10^2 < A < 10^{50}$ could survive.

\vspace{5mm} \noindent {\it PACS:}
26.60.Kp    % Equations of state of neutron-star matter
26.50.+x,   % Nuclear physics aspects of supernovae and other explosive environments

\end{abstract}

\begin{keyword}
Equation of state; Quark nuggets
\end{keyword}

\end{frontmatter}

%________________________________________________________________
\section{Introduction}           %% first-level sections will be auto-capitalized
\label{sect:intro}

It has long been proposed that much of the baryon number ($A$) of the Universe is condensed into the quark phase (usually called quark nuggets, QNs) during the quark-hadron phase transition~\cite{Witten84}. To survive in the hot QCD medium ($\sim$ 150 MeV), a QN of a certain size must outlive two decay processes, namely surface evaporation~\cite{evapor85} and boiling [nucleation of hadronic bubbles (HBs)]~\cite{boil89,boil91}. The former is generally very efficient when the environment is transparent to neutrinos, and the details mainly depend~\cite{evapor85} on the dynamic properties of the neutrino-driven cooling, for example the neutrino opacity. Our interest lies in the latter case, i.e., the boiling of QNs into hadrons, which is closely related to the underlying microscopic physics of the quark-hadron phase transition.

In one of the earliest studies, Alcock \& Olinto~\cite{boil89} described nucleons in terms of an ideal gas, and assumed
that the pressure in the strange-quark matter would be contributed entirely by the thermal spectrum of light particles (electrons, neutrinos, and photons). Based on the idea that if the total surface area of HBs exceeds the QN surface area, boiling would be inefficient, they found a baryon number minimum $A_{\rm boil}$ above which boiling is important, and concluded that this limit must be as high as $10^{46}$ - $10^{49}$. They have furthermore treated the surface tension of QNs, $\sigma$, as a free parameter and obtained
for it an unusually large lower limit, namely (178 MeV)$^3$, which would mean that almost all QNs could not survive boiling. Later Madsen \& Olesen~\cite{boil91} treated the hadron phase as a Walecka-type interacting neutron-proton-electron ($npe$) gas and also introduced the fermion pressure in the quark phase using the MIT bag-like model~\cite{mit84}. They found a rapid dependence of $A_{\rm boil}$ on the parameters ($\sigma, B$), where $B$ is the bag constant. They argued that QNs may survive boiling for some choice of ($\sigma, B$), and that therefore for such a case boiling is not the dominant decay process for QNs, compared to the evaporation mentioned above. In the recent work of Lugones \& Horvath~\cite{boil04}, quark pairing and the curvature energy were introduced in the quark phase and it was concluded that both boiling and surface evaporation would be suppressed by the pairing gap. The authors also argued that boiling might be unlikely for intermediate temperatures ($T < T_{\Delta} \sim 0.57\Delta$), where $\Delta$ is the pairing gap.

Clearly, the most important aspect for the boiling problem is how to treat the strong interaction between quarks (in the quark phase) and between hadrons (in the hadron phase). This will evidently affect the chemical composition of the two phases, and lead to different conclusions of the occurrence of boiling.
The aim of this work is hence to employ the hadronic and quark EOSs based on the most advanced microscopic approaches. The results should have important impacts on the conclusions reached before~\cite{boil91} based on the phenomenological nuclear many-body theory. We can also get more insight from the comparison of the calculated results using different versions of baryon interactions, that is, to achieve a better understanding on the relation between the cosmological QCD phase transition and the underlying EOS.

For the quark phase, the simple MIT bag model~\cite{mit84} used in previous studies~\cite{boil91,boil04} is actually not well justified, since it includes no interactions between quarks (quarks are asymptotically free within a large bag). Also, the model itself was originally proposed in order to treat quark confinement, therefore at any finite temperature, a more self-consistent scheme to treat thermal radiation and particle-antiparticle creation is needed. In the present study this is achieved by a fully self-consistent thermodynamic treatment of confinement, i.e., a recently developed quark mass density-dependent (QMDD) model~\cite{Peng00,Wen05}. This model has been widely used in the last few years for the structures and the viscosity of compact stars~\cite{liang08,Lug03,Zhe04,liangmn,liangraa}.

On the other hand, thanks to the rapid progress in the treatment of microscopic theories of the nuclear matter equation of state (EOS) in recent years, a detailed study of the hadron phase is by now possible, and we in the present work therefore can treat it much more accurately than in the phenomenological relativistic mean-field model used before~\cite{boil91}. We employ the parameter-free microscopic Brueckner-Hartree-Fock (BHF) approach that has been widely used for the study of dense stellar matter and neutron star properties~\cite{liang08,liang06,book,baldo,liang11,bbb,bhf02,bhf03,liang04,bhf06,liang10,bhfreport,bhf13aal,bhf13}, along with two cases of baryon interactions as inputs. They have the same nucleonic two-body potentials, Argonne v18~\cite{v18}, but different three-body forces (TBF), i.e., the phenomenological Urbana model~\cite{pheno95,pheno97}, and a microscopic TBF constructed from the meson-exchange current approach~\cite{micro89}. Both of them reproduce fairly well the saturation point of symmetric nuclear matter around the saturation density of $0.17$ fm$^{-3}$, and also fulfill the recent 2-solar-mass neutron star mass measurement~\cite{2nat,2sci}. They, however, give a very different
high-density EOS ($>0.4$ fm$^{-3}$)~\cite{liang06}. In particular, the microscopic TBF turns out to be more repulsive than the Urbana model at high densities, and the discrepancy between the two predictions becomes increasingly large as the density increases. Since the threshold of the quark-hadron transition is essentially determined by the stiffness of different hadron EOSs, we should keep in mind that the EOS from the microscopic TBF is stiffer than that of the phenomenological one. Hereafter, we refer to ``stiff EOS'' as
the one with the microscopic TBF, and to ``soft EOS'' as the one with the phenomenological TBF. Calculations are mainly done using the microscopic TBF, and results with the phenomenological TBF are presented as well in several cases for comparison.

Here we have neglected the possible appearance of hyperons and pion or kaon condensates in the hadron phase, which in general might soften the high-density EOS. How to confront them with the high-mass neutron stars is an important topic discussed frequently in recent papers~\cite{liang11,liangap,hyperonapj}. It would be straightforward to include the strangeness in the hadron phase in a subsequent study, once the controversial high-density EOS is clarified.

Let us also mention here that the subject of study of the boiling of QNs (into hadrons), in principle demands that the QCD phase transition is of first order (see for instance \cite{boil04} and references therein). Lattice QCD studies over the past years have however reached the conclusion that, for physical quark masses and a vanishing baryon chemical potential $\mu$, this transition is rather a smooth crossover than a first order phase transition \cite{Brown1990,Aoki1999,Aoki2006}.
If the Universe follows the ``standard" scenario and undergoes the QCD phase transition with only a very small $\mu$,
this would mean that QNs could not have been created and the discussion of their properties would thus be irrelevant for
the Universe that we live in. It should however be stressed here that there is room for an alternative scenario, which has been discussed in the literature~\cite{Kaempfer1986,Borghini2000,Boeckel2010,Boeckel2012}
as little inflation. In this case, the Universe follows a path with larger $\mu$ and can therefore undergo a first order phase transition as the QCD phase diagram is expected to have a critical endpoint at some finite value of $\mu$, above which the the quark-gluon plasma and hadron gas phases are separated by a first order phase transition line. The creation of QNs can hence not be ruled out and studying their properties may still be of
relevance for the physics of our Universe.

The paper is organized as follows. In section 2, we establish our physical model and describe in details the numerical methods for the calculation. In section 3, numerical results are discussed. We present our main conclusions in section 4.

\section{{\bf The} Model}
\label{sect:Mod}

%%%%%%%%%%%%%%%%%%%%%%%%%%%%%%%%%%%%%%%%%%%
\subsection{Boiling of QNs}

In the hot QCD medium, the hadron gas may be energetically favored in thermal fluctuations, and bubbles of hadronic gas would nucleate throughout the volume of the produced nuggets of strange matter. This process is called ``boiling of QNs''. If boiling happens, the QN would dissolute into hadrons and disappear in the Universe.

Following the estimation using classical nucleation theory by Alcock \& Olinto~\cite{boil89}, the work done to form a bubble
of radius $r$ composed by the hadronic phase inside the quark phase is
\begin{equation}
W = - \frac{4}{3} \pi r^3 \Delta P  + 4 \pi  \sigma  r^2, \label{work}
\end{equation}
where $\sigma$ is the QN surface tension. We follow Madsen \& Olesen~\cite{boil91} and self-consistently calculate the surface tension from all
fermion species ($i=u,d,s,e$) as:
\begin{equation}
\sigma_i = \frac{3T}{8\pi}\int_0^{\infty}(1-\frac{2}{\pi}\arctan\frac{k}{m_i})\ln[1+\exp(-\frac{e_i(k)-\mu_i}{T})]kdk.
\end{equation}
where the energy $e(k)$ is given as
$e(k) = \sqrt{k^2+m_i^2}$ with $m_i$ ($\mu_i$) being the mass (chemical potential) of component
$i$. $T$ is the temperature.
$\Delta P = P_H - P_Q$ is the pressure difference between the hadron phase (with a pressure of $P_H$) and the quark phase (with a pressure of $P_Q$). Assuming that the phase transition is first order, its properties are
calculated from the the pressure difference between the two phases based on the chemical equilibrium condition:
\begin{equation}
\mu_Q(P_Q) = \mu_H(P_H) \equiv \mu \label{mu}
\end{equation}
where $\mu_Q $ and $\mu_H$ are the baryon chemical potentials for the hadron and quark phases, respectively.

\begin{figure}
\begin{center}
\includegraphics[width=12cm,clip=,angle=0]{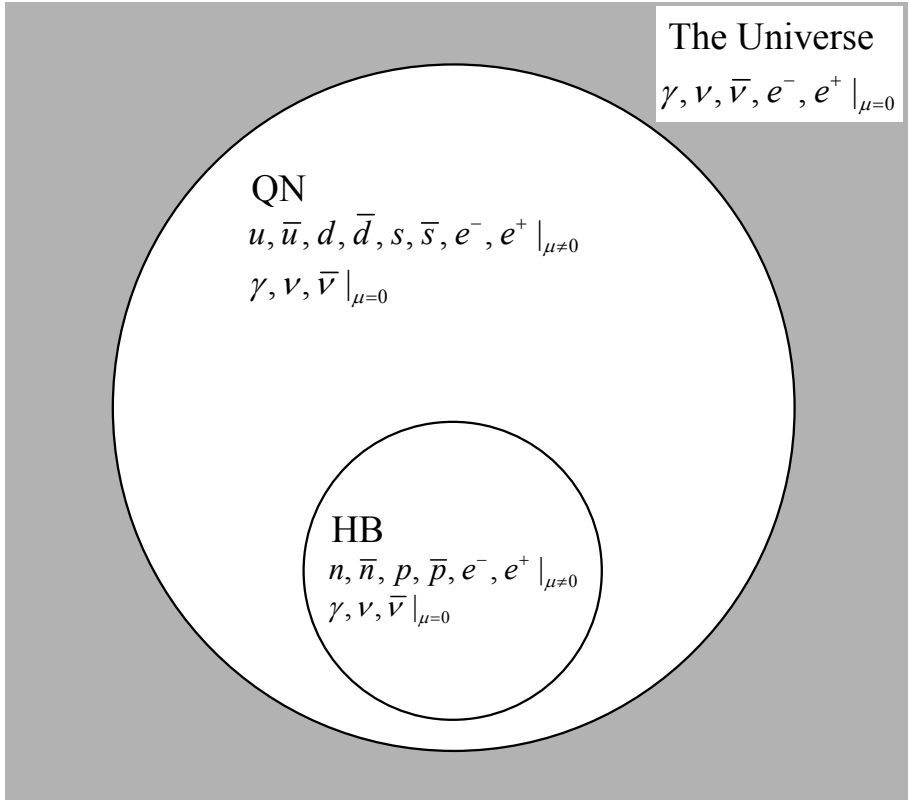}
\end{center}
\caption{Pictorial representation of the model. Zero chemical potential ($\mu=0$) represents that the relevant particles are thermally produced.\label{fig1}}
\end{figure}

Specifically, as illustrated in Fig.~\ref{fig1}, besides the common pressures from thermal photons and neutrinos, $P_H$ is contributed by hadrons and $e^{\pm}$ pairs which will be dealt as accurately as possible here, $P_Q$ constitutes a nonthermal pressure from $u,d,s$ quarks and $e^{\pm}$ pairs. Since the pressure of the quark phase, $P_Q$, is equal to the pressure in the Universe (mainly contributed by thermal photons, neutrinos, electrons, and positrons), the nonthermal contributions of quarks and $e^{\pm}$ pairs in the quark phase balance
exactly the thermal electrons and positrons in the Universe.
This constrain is used to determine the independent baryon chemical potential, $\mu$. Clearly, $\Delta P$ comes from the pressure difference apart from the common contribution from thermal photons and neutrinos,
and can be computed from the pressure of hadron phase, $P_H$, subtracted by the pressure from thermally produced electrons and positrons:
\begin{equation}
\Delta P = P_H -  \frac{8 \pi}{3(2 \pi \hbar)^3}
 \int_0^{\infty}(\mbox{f}_{e^-}+\mbox{f}_{e^+})  k^4 /e(k)d k |_{\mu_e=0}
 \label{dp}
\end{equation}
with $\mbox{f}_i$ the Fermi--Dirac distribution
written as
\begin{eqnarray}
\mbox{f}_{e^-}(k,T) & = & \frac{1} {{\rm
exp}[(e(k) - \mu_e)/k_BT]+1}, \\
\mbox{f}_{e^+}(k,T) & = & \frac{1} {{\rm exp}[(e(k) + \mu_e)/k_BT]+1}, \label{fermi}
\end{eqnarray}
where $k_B$ is the Boltzmann constant and the energy $e(k) = \sqrt{k^2+m_e^2}$ with $m_e$ ($\mu_e$) being the mass (chemical potential) of the electrons or positrons.

Whether boiling is important or not depends on the formation rate of critical bubbles, since only those bubbles with a radius greater than the critical radius will be able to grow. By maximizing $W$ in Eq.~(\ref{work}) w.r.t. $r$, the work of a critical-size bubble can be obtained as
\begin{equation}
W_c = \frac{16\pi}{3}\frac{\sigma^3}{\Delta P^2}. \label{W_c}
\end{equation} 
The rate at which critical bubbles appear is then
\begin{equation}
p(T,\mu)\sim T^4 \exp(-W_c/T).
\end{equation}
If the number of bubbles becomes so small that the total surface area
of the bubbles is smaller than the bounding surface area
of a QN, boiling would be inefficient. That gives a
minimum baryon number $A_{\rm boil}$ above which boiling is
efficient~\cite{boil91}:
\begin{equation}
A_{\rm boil}\approx 7.9 \times 10^{-61} \frac{\Delta P^6}{T^6\sigma^6}\exp(16\pi\frac{\sigma^3}{T\Delta P^2}). \label{boil}
\end{equation}
Much of the microscopic physics introduced in the present work, including the quark-quark interaction, would enter in $\Delta P$ and consequently influence the physics of boiling. As will be shown later, the high powers and exponential term in Eq.~(\ref{boil}) would result in a rapid dependence on parameters ($\Delta P$, $\sigma$, $T$).

\begin{figure}
\begin{center}

\includegraphics[width=0.45\textwidth]{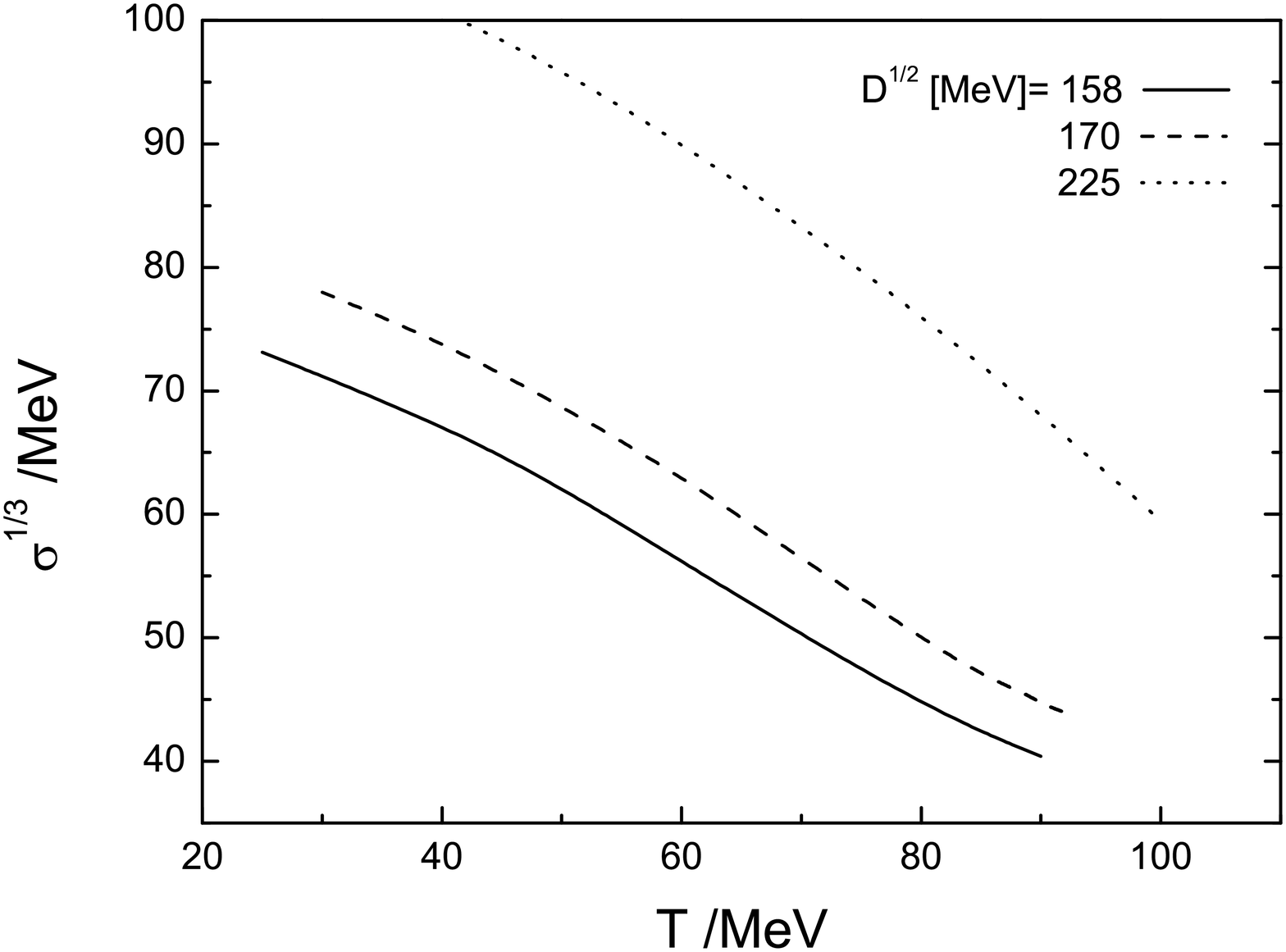}
\includegraphics[width=0.45\textwidth]{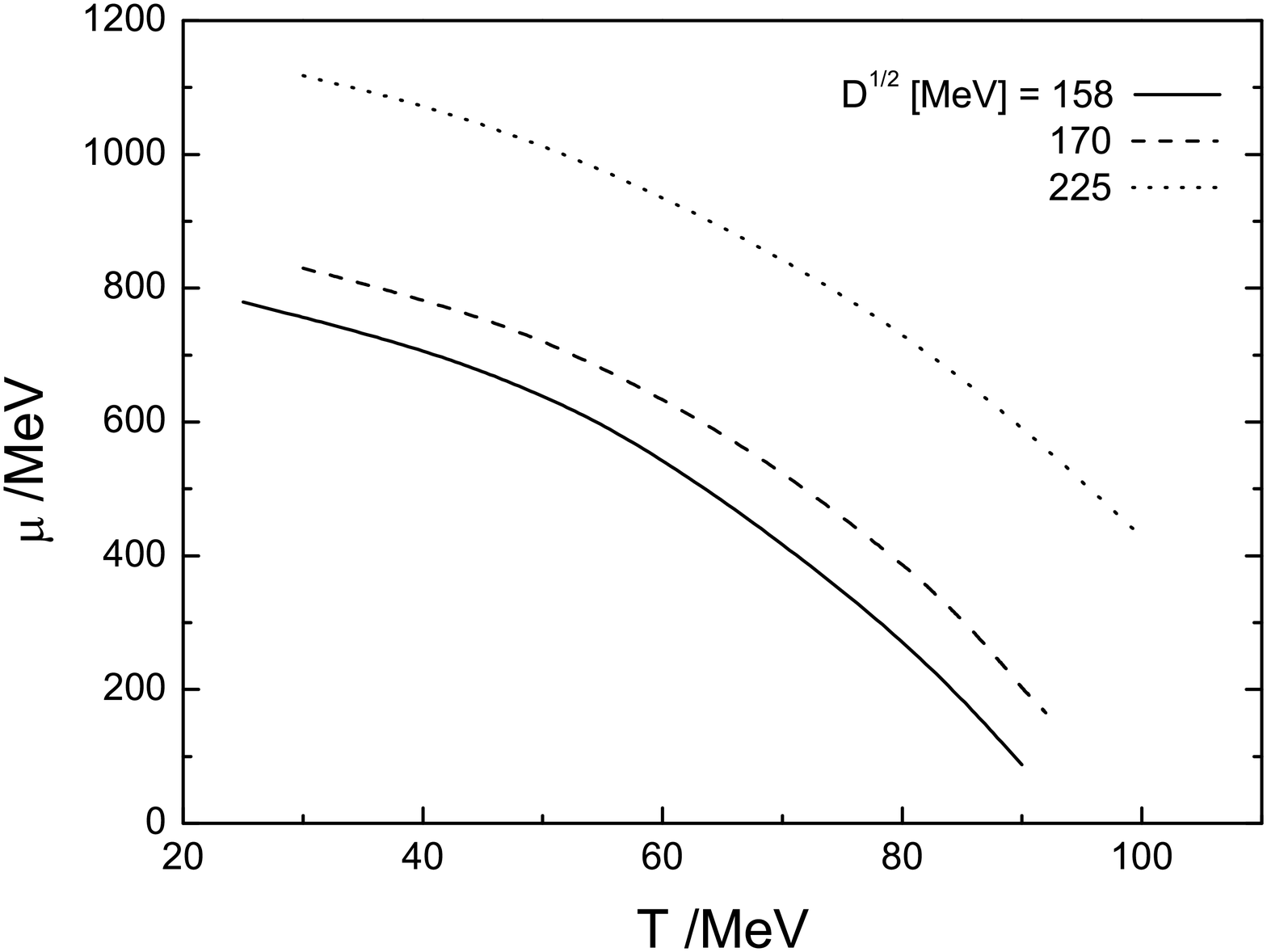}
\end{center}
\caption{(Left) surface tensions $\sigma$ and (right) baryon chemical potentials $\mu$ are shown as a function of temperature $T$ for three values of the quark confining parameter $D^{1/2} = 158$ MeV, $170$ MeV, $225$ MeV in the quark phase. The microscopic TBF is employed for the hadron phase.\label{fig2}}
\end{figure}

The remaining part of this section is devoted to the calculation of $\Delta P$.
As mentioned above, assuming the phase transition is first order, calculations can be done separately for two phases. They are related by the
constraint of the common chemical potentials $\mu$ of Eq.~(\ref{mu}). In the following, theoretical models for the the hadron phase and the quark phase are illustrated, respectively.

\subsection{The hadron phase}

Let us first study the hadron phase, that is nuclear matter consisting of nucleons and $e^{\pm}$ pairs in
$\beta$-equilibrium of the following weak reactions:
\begin{eqnarray}
n     \rightleftharpoons  p+e^- + \bar{\nu}_e, \\
 e^+ + e^- \rightleftharpoons \nu_{e} +
\bar{\nu}_{e}
\end{eqnarray}
Under the condition of neutrino escape, this equilibrium can be written as
\begin{eqnarray}
\mu_n - \mu_p = \mu_{e^-} = -\mu_{e^+}. \label{mu_N}
\end{eqnarray}
The requirement of charge neutrality implies
\begin{eqnarray}
n_p = n_{e^-} - n_{e^+}, \label{neutral}
\end{eqnarray}
where $n_i$ is the number density of component
$i$.

The chemical potentials of the non-interacting leptons $e^{\pm}$ are obtained by solving numerically the free Fermi gas model at
finite temperature. The nucleonic chemical potentials required in Eq.~(\ref{mu_N}) are derived from the free energy density of
nuclear matter, based on the finite-temperature BHF nuclear many-body approach discussed below.

The BHF approach~\cite{book} is one of the most advanced microscopic approaches to the EOS of
nuclear matter. Recently, this model was extended to the finite-temperature regime within the
Bloch-De Dominicis formalism \cite{bloch1,bloch2,bloch3}. The
central quantity of the BHF formalism is the $G$-matrix, which in
the finite-temperature extension
\cite{book,baldo,bloch1,bloch2,bloch3} is determined by solving
numerically the Bethe-Goldstone equation, and can be written in
operatorial form as \beq
  G_{ab}[W] = V_{ab} +
  \sum_c \sum_{p,p'}
  V_{ac} \big|pp'\big\rangle
  { Q_c \over W - E_c +i\eps}
  \big\langle pp'\big| G_{cb}[W]\:,
\label{e:g} \eeq where the indices $a,b,c$ indicate pairs of
nucleons and the Pauli operator $Q$ and energy $E$ determine the
propagation of intermediate nucleon pairs. In a given
nucleon-nucleon channel $c=(12)$ one has \beq
   Q_{(12)} &=& [1-\mbox{f}_1(k_1)][1-\mbox{f}_2(k_2)]\:,
\eeq \beq
  E_{(12)} &=& m_1 + m_2 + e_1(k_1) + e_2(k_2)\:,
\label{e:e} \eeq with the single-particle (s.p.) energy $e_i(k) =
k^2\!/2m_i + U_i(k)$, the Dirac-Fermi
distribution
$\mbox{f}_i(k)=\big( e^{[e_i(k) - \tilde{\mu_i}]/T} + 1 \big)^{-1}$, the
starting energy $W$, and the above-mentioned baryon interaction
$V$ as fundamental input. The various
single-particle (s.p.)
potentials within the
continuous choice are given by \beq
  U_1(k_1) = {\rm Re}\!\!\!\!
  \sum_{2=n,p}\sum_{k_2} n(k_2)
  \big\langle k_1 k_2 \big| G_{(12)(12)}\left[E_{(12)}\right]
  \big| k_1 k_2 \big\rangle_A\:,
\label{e:u} \eeq where $k_i$ generally denote momentum and spin. For
given partial densities $n_i\; (i=n,p)$ and temperature $T$,
Eqs.~(\ref{e:g}-\ref{e:u}) have to be solved self-consistently along
with the equations for the auxiliary chemical potentials
$\tilde{\mu_i}$, $ n_i = \int_k \mbox{f}_i(k) $.

Once the different s.p.~potentials for the species $i=n,~p$ are
known, the free energy density of nuclear matter
can be obtained using
the following simplified expression \beq
 f_N = \sum_i \left[ \sum_{k} \mbox{f}_i(k)
 \left( {k^2\over 2m_i} + {1\over 2}U_i(k) \right) - Ts_i \right]\:,
\label{e:f} \eeq where \beq
 s_i = - \sum_{k} \Big(\mbox{f}_i(k) \ln \mbox{f}_i(k) + [1-\mbox{f}_i(k)] \ln [1-\mbox{f}_i(k)] \Big)
\eeq is the entropy density for component $i$ treated as a free gas
with s.p.~spectrum $e_i(k)$~\cite{baldo,book}. All thermodynamic quantities of interest can then be computed from
the free energy density, Eq.~(\ref{e:f}); namely, the ``true"
chemical potentials $\mu_i\; (i=n,~p)$, internal energy density
$\eps_N$, and pressure $P_N$ are \beq
 \mu_i &=& {{\partial f_N}\over{\partial n_i}} \:,
\\
s_N &=& -{{\partial f_N}\over{\partial T}} \:,
\\
 \eps_N &=& f_N + Ts_N \:,
 \\
  P_N &=& n_B^2 {\partial{(f_H/n_B)}\over \partial{n_B}}
 = \sum_i \mu_i n_i - f_N\:.
\label{e:eps} \eeq

One can then proceed to calculate the composition of the hot $\beta$-equilibrium matter by solving Eqs. (\ref{mu_N}) and
(\ref{neutral}), together with the conservation of the baryon number, $n_n+n_p = n_B$. Then the
total energy density $\epsilon$ and the total pressure $P$ of the system
are
\begin{eqnarray}
 \eps_H &=& \eps_l + \epsilon_N\:,  \eeq \beq
 P_H &=& P_l + P_N\:,
\end{eqnarray}
where $\epsilon_l$ and $P_l$ are the standard contributions of the leptons:
 \beq
 \eps_{l}  & = & \frac{8 \pi}{(2 \pi \hbar)^3}
 \int_0^{\infty}(\mbox{f}_{e^-}+\mbox{f}_{e^+})  e(k) k^2 d k\:,
   \\
 P_{l} & = &\ \frac{8 \pi}{3(2 \pi \hbar)^3}
 \int_0^{\infty}(\mbox{f}_{e^-}+\mbox{f}_{e^+})  k^4 /e(k)d k\:.
 \label{eq:electroneps} \
\eeq

\subsection{The quark phase}

We consider the quark phase as a mixture of interacting $u$,
$d$, $s$ quarks, electrons and positrons. In the QMDD model, the mass of the quarks $m_i$
($i = u, d, s$)  is parameterized
by the baryon number density
$n_B$ as follows
\begin{eqnarray}
m_i \equiv m_{i0}+ m_{\mathrm{I}}=m_{i0}+\frac{D}{n_B^z},
\label{mqT0}
\end{eqnarray}
The density-dependent mass $m_{i}$ includes two parts: one is the
original mass or current mass $m_{i0}$, the other is the interacting
part $m_{\mathrm{I}}$. The light-quark masses are very small, and we simply take $m_{u0}=m_{d0}=0$. As to the uncertain strange quark mass, a modest value of $m_{s0}=95$ MeV is chosen. In principle, the quark mass scaling should
be determined from QCD, which is obviously impossible presently. As mentioned before, we use the cubic scaling $z$ = 1/3 \cite{Pen99}, based on the linear confinement and in-medium chiral condensates. It was first derived at zero temperature \cite{Pen99} and then expanded to finite temperature \cite{Wen05}.

$D$ is the confinement parameter. From our previous work of Peng et al. 2008, we know that
it has a lower bound of $D^{1/2} = 156$ MeV, and an upper bound of $D^{1/2} = 270$ MeV.
The lower bound comes from the nuclear physics constraint, demanding that at $P=0$,
non-strange nuclear matter should be stable against decay to $(ud)$ quark matter.
This leads to the condition $E/A > M_{^{56}Fe}c^2/56 = 930$ MeV for $(ud)$ quark matter,
which gives the above-mentioned lower bound. The
upper bound can be derived from a relation between $D$ and the quark-condensate~\cite{liang08} and the known range of values for this condensate. The upper boundary of $270$ MeV is in fact a very conservative one. According to the updated quark condensate determined nowadays very precisely by lattice QCD~\cite{upqcd}, a range of ($161$ MeV, $195$ MeV) can be obtained.
Therefore in this work, we take three typical values of the confinement parameter as $D^{1/2} = 158$ MeV, $170$ MeV, $225$ MeV. The lower one ($158$ MeV) is chosen to account for the realizable case of absolutely stable strange-quark matter~\cite{liang08}, the middle one ($170$ MeV) because it satisfies the constraints from the newest lattice QCD results~\cite{upqcd}, and the large one ($225$ MeV) to study the situation in the region of the upper boundary.

The relevant chemical potentials $\mu_u$, $\mu_d$, $\mu_s$, and
$\mu_{e^{\pm}}$ satisfy the weak-equilibrium condition (we again assume that
neutrinos leave the system freely):
\begin{eqnarray}
 \mu_u- \mu_d = \mu_{e^-} = -\mu_{e^+}, ~~~\mu_d=\mu_s{\bf .} \label{weak}
\end{eqnarray}
The baryon number density and the charge density can be given as
\begin{eqnarray} \label{qmeq3}
n_B=\frac{1}{3}(n_u+n_d+n_s){\bf ,}
\end{eqnarray}
\begin{eqnarray} \label{qmeq4}
q_Q=\frac{2}{3}n_u-\frac{1}{3}n_d-\frac{1}{3}n_s+n_{e^+}-n_{e^-}.
\end{eqnarray}
The charge-neutrality condition requires $q_Q=0$.

In the present model, the single-particle energies depend on density and temperature via the quark masses. Based on the quasiparticle
assumption, the quark energy density can be written as ($i = u, d, s$)~\cite{liang08}
\begin{eqnarray}
\eps_Q &=& g\sum_i \sum_{k} \sqrt{k^2+m_i^2}\,\mbox{f}_i(k,T).
\label{ass1}
\end{eqnarray}
where the statistical weight
is
$g=6$ for quarks.
From the Landau definition of the single-particle energy extended
to finite temperature, we have
\begin{eqnarray}
\varepsilon_i(k)
 &=& \frac{\delta \eps_Q}{\delta \mbox{f}_i(k,T)}
 \nonumber \\
 &=&
    \sqrt{k^2+m_i^2}
     +g\sum_j \frac{m_j \mbox{f}_j(k,T)}{\sqrt{k^2+m_j^2}}
      \frac{\partial m_j}{\partial n_i}
 \nonumber \\
&\equiv& e_i(k)-\mu_{\mathrm{I}},
\end{eqnarray}
where $e_i(k)\equiv\sqrt{k^2+m_i^2}$\ is the usual dispersion
relation of free particles.
The extra term $\mu_{\mathrm{I}}$ can be added to the chemical
potential, so defining
\begin{eqnarray}
\mu_i^*\equiv\mu_i+\mu_{\mathrm{I}}.
\end{eqnarray}
Accordingly, the net density of the particle type $i$ is
$
n_i=g\sum_{k} \left[\mbox{f}_i(k,T)
    -\mbox{f}_{\bar{i}}(k,T)\right],$\
or, explicitly, we have
\begin{eqnarray}
n_i
= g\int_0^{\infty}
  \left\{\frac{1}{1+e^{[\varepsilon_i(k)-\mu^*_i]/T}}\right.
  \left.-\frac{1}{1+e^{[\varepsilon_i(k)+\mu^*_i]/T}}\right\}
  \frac{p^2\mbox{d}k}{2\pi^2}.
\end{eqnarray}
Inverting this equation, one determines $\mu^*_i$ as a function
of $n_i$ so that the free energy density of quarks can be given as
\begin{eqnarray}
f_q
=\sum_i
 \left[f_i^++f_i^-\right]
\label{FTtot}
\end{eqnarray}
with
\begin{eqnarray}
f_i^{\pm}
&=&
 g
 \int_0^{\infty}
 \bigg\{
 -T\ln\left[1+e^{-(\sqrt{k^2+m_i^2}\mp\mu_i^*)/T}
    \right]
\nonumber\\
&& \phantom{g_i\int_0^{\infty}\bigg\{}
 \pm\frac{\mu_i^*}{1+e^{(\sqrt{k^2+m_i^2}\mp\mu_i^*)/T}}
 \bigg\} \frac{k^2\mbox{d}k}{2\pi^2}.
 \label{FiTmum}
\end{eqnarray}
One can then determine the real chemical
potentials and pressure, according to the well-known relations ($i=u,d,s$):
\begin{eqnarray}
\mu_i = \frac{\partial f_q}{\partial n_i}, \ \
P_q     =-f_q+\sum_i\mu_in_i.
\end{eqnarray}
Solving Eqs. (\ref{weak}), (\ref{qmeq3}) and (\ref{qmeq4}), the total pressure of the system can be obtained:
\begin{eqnarray}
 P_Q &=& P_l + P_q\:,
\end{eqnarray}
after adding the contribution of the leptons $P_l$ (Eq.~(\ref{eq:electroneps})).

\section{Results and Discussion}
\label{sect:Res}

\begin{figure}
\begin{center}
\includegraphics[width=12cm,clip=,angle=0]{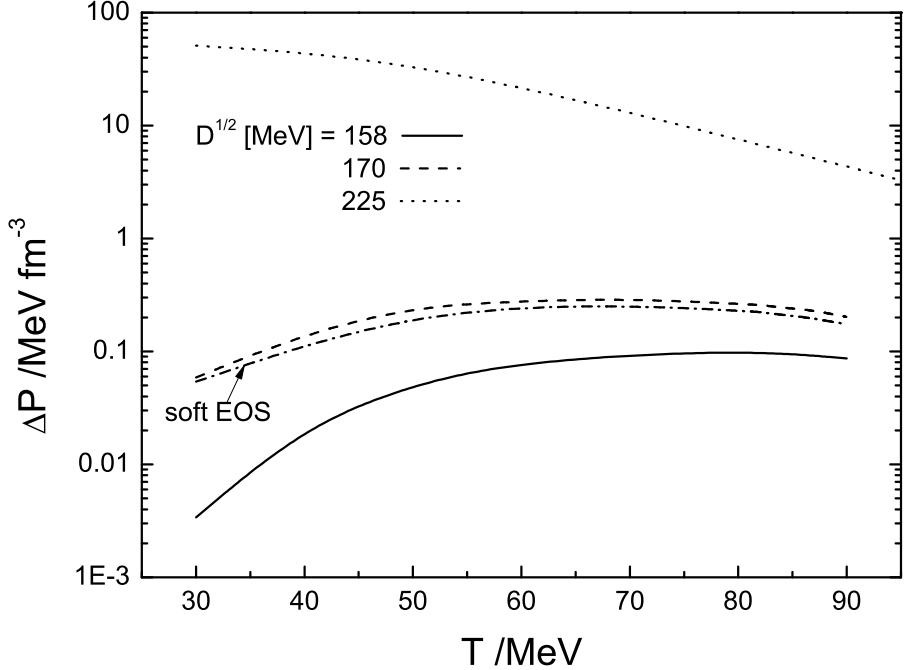}
\end{center}
\caption{Pressure differences $\Delta P$ are shown as a function of temperature $T$ for three values of the quark confining parameter $D^{1/2} = 158$ MeV, $170$ MeV, $225$ MeV in the quark phase. The microscopic TBF is employed for the hadron phase. Results from the the phenomenological TBF (labelled as ``soft EOS'') is also shown for the case of $D^{1/2} = 170$ MeV.}
\label{fig4}
\end{figure}

We begin in Fig.~\ref{fig2} with the surface tensions of QNs (left) and the baryon chemical potentials (right) as a function of temperature for three values of the quark confining parameter in the quark phase, and the microscopic TBF chosen for the hadron phase. During the cooling of the Universe, $\sigma$ increases, which means that QNs tend to be more bound. Also,
the larger the confining parameter, the larger the surface tension, which is reasonable because quarks interact more strongly. The baryon chemical potentials are decreasing functions with the temperature. With the increase of the confining parameter $D$, the equilibrium chemical potential $\mu$ of the two phases increases due to a larger pressure difference $\Delta P$ as shown in Fig.~\ref{fig4}.
%\begin{figure}
%\begin{center}
%\includegraphics[width=13cm,clip=,angle=0]{fig3.eps}
%\end{center}
%\caption{Baryon chemical potentials $\mu$ are shown as a function of temperature $T$ for three values of the quark confining parameter $D^{1/2} = 158$ MeV, $170$ MeV, $225$ MeV in the quark phase. The microscopic TBF is employed for the hadron phase.}
%\label{fig3}
%\end{figure}

%Fig.~\ref{fig3} shows the baryon chemical potentials as a function of temperature for three values of the quark confining parameter in the quark phase, and the microscopic TBF chosen for the hadron phase.
%They
To see the influence of the hadron EOS, results from the the phenomenological TBF (labelled as ``soft EOS'') is also shown for the case of $D^{1/2} = 170$ MeV. It is clearly
observed that a soft hadron EOS will result in a slight decrease of the pressure difference of the two phases. It has a similar decreasing effect on the equilibrium chemical potential $\mu$, but the difference is very small and can not be distinguished in Fig.~\ref{fig2}.

\begin{figure}
\begin{center}
\includegraphics[width=12cm,clip=,angle=0]{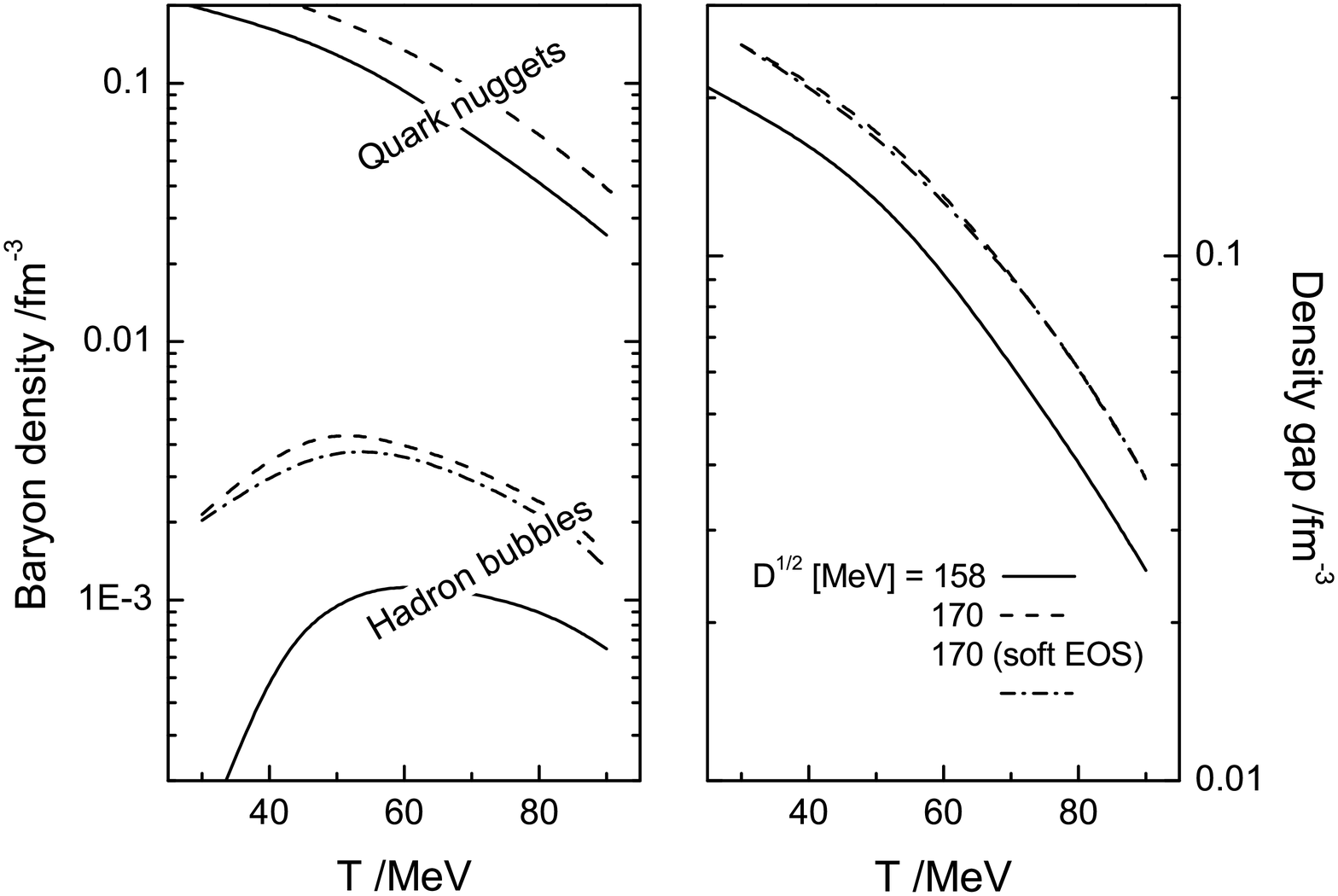}
\end{center}
\caption{Baryon densities of the two phases (left panel) and density gaps between the two phases (right panel) at chemical equilibrium are shown as a function of the temperature $T$, for two values of the quark confining parameter $D^{1/2} = 158$ MeV (solid lines) and $170$ MeV (dashed lines) in the quark phase. The microscopic TBF is employed for the hadron phase. Results from the the phenomenological TBF (labelled as ``soft EOS'') is also shown for the case of $D^{1/2} = 170$ MeV (dash-dotted lines).}
\label{fig5}
\end{figure}

More microscopically, we show in the left panel of Fig.~\ref{fig5} the baryon number densities of the two phases at chemical equilibrium. The densities in QNs always decrease with the temperature $T$, but the densities in HBs first increase with $T$ then decrease with it. Soft EOS in the hadron phase will in general lower
the baryon densities of HBs at chemical equilibrium
only slightly for each temperature $T$.
Compared with that of the hadron EOS, the effect of
the quark confining parameter $D$ is more evident.
A larger $D$ value will result in increasing baryon densities both in QNs and HBs, and also lead to a
larger density gap
between the two phases, as shown in the right panel of Fig.~\ref{fig5}. One can also notice a small decreasing effect of the soft hadron EOS on the density gap.

\begin{figure}
\begin{center}
\includegraphics[width=12cm,clip=,angle=0]{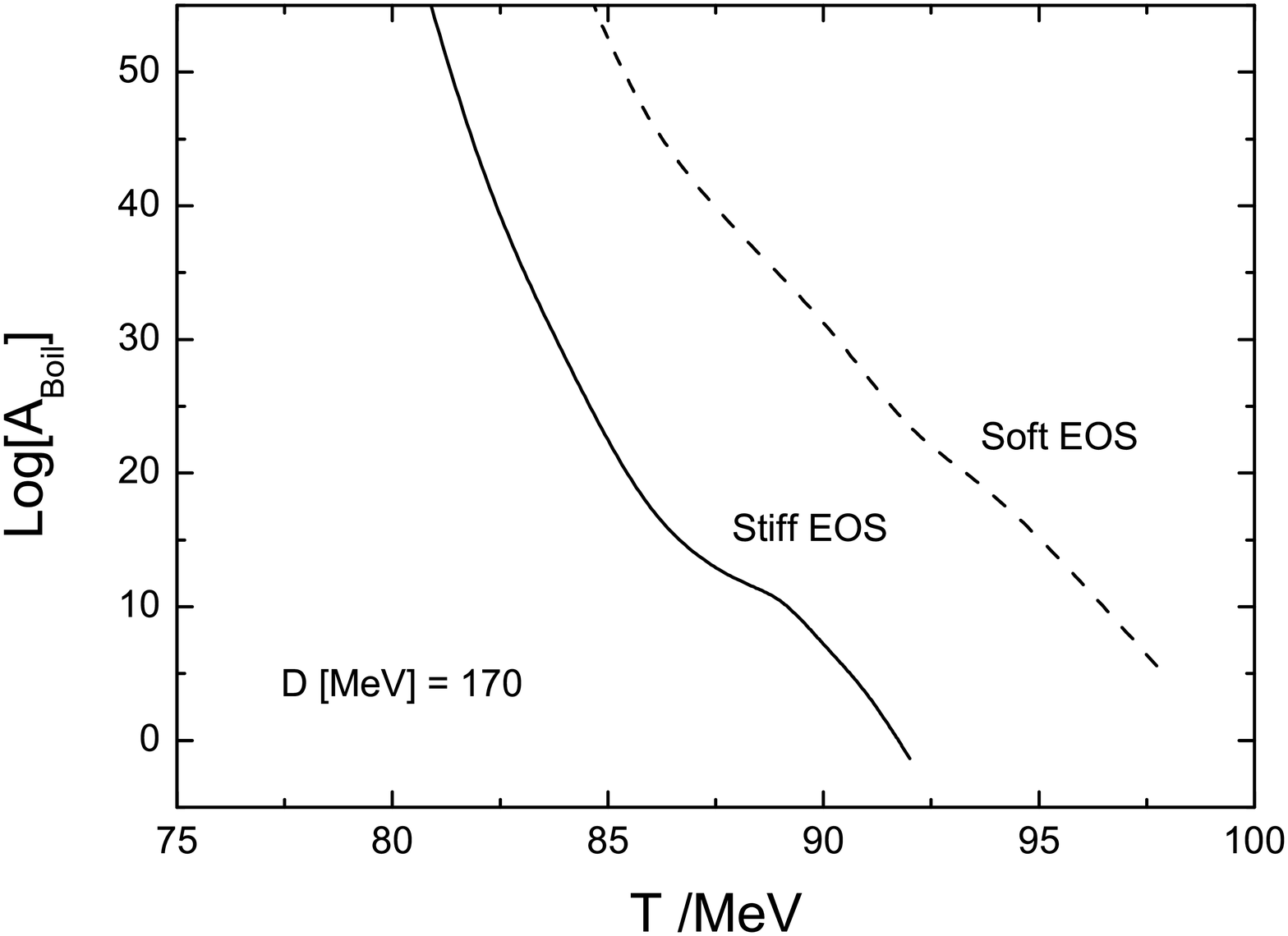}
\end{center}
\caption{Baryon number limits $A_{\rm boil}$, above which boiling could happen, are shown as a function of the temperature $T$. The calculation is done with $D^{1/2} = 170$ MeV in the quark phase, for both the microscopic TBF (solid line, labelled as ``stiff EOS'') and the phenomenological TBF (dashed line, labelled as ``soft EOS'') for the hadron phase. The cases of $D^{1/2} = 158$ MeV and $D^{1/2} = 225$ MeV would result in $A_{\rm boil}$ values far from the baryon number of a cosmological QN, therefore they are not shown here.}
\label{fig6}
\end{figure}

For a smaller density gap between the two phases, the boiling of a QN to HBs becomes more difficult, as it is clearly demonstrated in Fig.~\ref{fig6}, where the baryon number limits $A_{\rm boil}$, above which boiling could happen, are shown as a function of the temperature $T$. The calculation is done with $D^{1/2} = 170$ MeV in the quark phase, for both the microscopic TBF and the phenomenological TBF for the hadron phase. A slightly smaller pressure difference (as shown in Fig.~\ref{fig4}), or a smaller density gap (as shown in Fig.~\ref{fig5}) in the soft EOS case (namely the case of the phenomenological TBF) will result in a larger work for a critical-size bubble, and finally lead to a lower probability of bubbles nucleation, namely a larger value of the limit $A_{\rm boil}$.

More importantly, it is found that the limit $A_{\rm boil}$ is a sharp decreasing function of the temperature $T$, as indicated by Eq.~(\ref{boil}).
With the chosen $D$ parameter, large amounts of cosmological quark nuggets between $10^2 < A < 10^{50}$ would boil around $T = 90$ MeV.
A small change of the $D$ value from 170 MeV would result in a sensitively sharp
increase
of the $A_{\rm boil}$ value, which is far from the baryon number of a cosmological QN. This is what happens in the cases of 158 MeV and 225 MeV. It means that only for a small parameter range of the confining parameter $D$, there is a chance that the QN boiling could happen very efficiently during the early stage ($T \sim $ 90 MeV) of the cooling. Otherwise, the baryon limit is irrelevant and no QNs could boil and might survive until the present time. We mention here that our calculation demonstrates an amazing coincidence of the lattice results~\cite{upqcd} just falling into the range where boiling could be important.

\section{Conclusions}

In this paper, we have renewed the study of the boiling of possible QNs during the quark-hadron phase transition of the Universe at nonzero chemical potential. For this purpose, a parameter-free microscopic BHF model extended to finite temperature is employed for the hadron phase to describe HBs, and a quark model with self-consistent thermodynamic treatment of the confinement is used to deal with the quark phase for the description of QNs. Both
phases are in beta-stable equilibrium through weak processes. The phase transition is regarded as a first-order one, in order to get the pressure difference between the two phases. Also, the important parameter of the QN surface tension is calculated self-consistently at each sets of parameters. Detailed presentations of the QN surface tension, the baryon chemical potential, as well as the pressures and the baryon densities for the two phases are shown.

We found a larger effect of the confinement parameter, than that of the hadron EOS. The baryon number limits are found above which boiling could be efficient. It turns out that only a limited range of the confinement parameter allow boiling to happen around a temperature of 90 MeV. This allowed range of the confinement parameter falls comfortably into the range
provided by lattice QCD. This therefore appears to be the favoured scenario at present.
Large numbers of QNs would boil to HNs in this case and QNs may not exist in the present Universe.
For other values of this parameter, boiling is impossible to happen, and QNs with a cosmological baryon number of $10^2 < A < 10^{50}$ could be possibly found. Future experiments of detecting dark matter can provide a crucial cross-check to this problem.

Furthermore, the importance of quark pairing gap was demonstrated in the work of Lugones \& Horvath~\cite{boil04} within the MIT model, taking into account the BCS-type of pairing for quarks. Therefore, a more proper treatment of pairing, for example using the Nambu-Jona-Lasinio (NJL), or the Polyakov-NJL model (PNJL), should be interesting. Since the NJL (or PNJL) model, not like the MIT + BCS scheme, incorporates all the symmetries known from the fundamental theory of strong interactions (i.e., quantum chromodynamics), it could improve our understanding of the boiling problem and its relation with the phase diagram of the high-density strong interaction matter.

\section{Acknowledgments}
Two of us (AL, TL) would like to thank Dr. H. Mao for his suggestions for the manuscript and his hospitality during our stay in Hangzhou. This work was supported by the National Basic Research Program (973 Program) of China Grant (Nos 2012CB821800, 2014CB845800), the Fundamental Research Funds for the Central Universities, the National Natural Science Foundation of China (Grant Nos 11078015, 11103015, 11225314, 11233006, U1331101), and the CAS Open Research Program of Key Laboratory for the Structure and Evolution of Celestial Objects under grant OP201305.

\end{document}